\documentclass[12pt,preprint]{aastex}

\cpright{AAS}{2011}
\shortauthors{Wang, Robbrecht, and Muglach}
\shorttitle{Dark Canopies Around Active Regions}

\begin{document}

\title{The Evolution of Dark Canopies Around Active Regions}
\author{Y.-M. Wang}
\affil{Code 7672W, Space Science Division, Naval Research Laboratory, 
Washington, DC 20375-5352, USA}
\email{yi.wang@nrl.navy.mil}
\author{E. Robbrecht}
\affil{Royal Observatory of Belgium, 1180 Brussels, Belgium}
\email{eva.robbrecht@oma.be}
\and
\author{K. Muglach\altaffilmark{1}}
\affil{Code 674, NASA Goddard Space Flight Center, Greenbelt, MD 20771, USA}
\email{karin.muglach@nasa.gov}
\altaffiltext{1}{Also at ARTEP, Inc., Ellicott City, MD 21042, USA.}

\begin{abstract}
As observed in spectral lines originating from the chromosphere, 
transition region, and low corona, active regions are surrounded by 
an extensive ``circumfacular'' area which is darker than the quiet Sun.  
We examine the properties of these dark moat- or canopy-like areas 
using \ion{Fe}{9} 17.1~nm images and line-of-sight magnetograms from the 
{\it Solar Dynamics Observatory}.  The 17.1~nm canopies consist of fibrils 
(horizontal fields containing EUV-absorbing chromospheric material) 
clumped into featherlike structures.  The dark fibrils initially form a 
quasiradial or vortical pattern as the low-lying field lines fanning out 
from the emerging active region connect to surrounding network and 
intranetwork elements of the opposite polarity.  The area occupied by 
the 17.1~nm fibrils expands as supergranular convection causes the 
active region flux to spread into the background medium; the outer boundary 
of the dark canopy stabilizes where the diffusing flux encounters a 
unipolar region of the opposite sign.  The dark fibrils tend to accumulate 
in regions of weak longitudinal field and to become rooted in 
mixed-polarity flux.  To explain the latter observation, we note 
that the low-lying fibrils are more likely to interact with small loops 
associated with weak, opposite-polarity flux elements in close proximity, 
than with high loops anchored inside strong unipolar network flux.  
As a result, the 17.1~nm fibrils gradually become concentrated around the 
large-scale polarity inversion lines (PILs), where most of the 
mixed-polarity flux is located.  Systematic flux cancellation, assisted by 
rotational shearing, removes the field component transverse to the PIL 
and causes the fibrils to coalesce into long PIL-aligned filaments.
\end{abstract}


\section{Introduction}
Active regions are often surrounded by an area whose brightness is reduced 
even relative to that of the quiet Sun.  This phenomenon was first noted 
in observations of ``circumfacular regions'' in the \ion{Ca}{2} K-line 
(Hale \& Ellerman 1903; St. John 1911).  The dark areas, also prominent 
in \ion{Ca}{2} 854.2~nm spectroheliograms, were subsequently shown 
to coincide approximately with the iron-filing or ``vortex'' pattern 
of H$\alpha$ fibrils that rapidly appears around emerging active regions 
(Howard \& Harvey 1964; Bumba \& Howard 1965; Veeder \& Zirin 1970; 
Foukal 1971a, 1971b; Harvey 2005, 2006; Rutten 2007; Cauzzi et al. 2008; 
Reardon et al. 2009).  During the first week of development, the major axis 
of the fibril pattern expands at a rate of $\sim$0.2~km~s$^{-1}$, 
corresponding to roughly one supergranular cell radius per day.

With the advent of UV and extreme-ultraviolet (EUV) observations, it is 
now evident that the reduced emission around active regions characterizes 
a wide range of spectral lines originating in the chromosphere, 
transition region, and low corona (Moses et al. 1997; Feldman et al. 2000).  
Indeed, in emission lines formed at temperatures $T\sim 0.5$--1.0~MK 
and in \ion{He}{2} 30.4~nm, these large moatlike areas, or 
``active-region canopies'' as we henceforth call them, are 
sometimes mistaken for coronal holes.  Like their H$\alpha$ counterparts, 
the EUV canopies consist of dark fibrilar structures, which are generally 
thought to trace out horizontal magnetic fields in the chromosphere.

Here, we use \ion{Fe}{9} 17.1 nm images and longitudinal magnetograms 
recorded with the {\it Solar Dynamics Observatory} ({\it SDO}) to study 
the properties of dark canopies around active regions.  Our main objective 
is to clarify the relationship between the 17.1~nm fibril structures and 
the evolving photospheric field, on the assumption that the fibrils are 
aligned with the chromospheric field.  We do not address the question of how 
the EUV canopies differ from those seen in the visible, or the related 
and more difficult question of how the spectral lines themselves are formed.

\section{Observations}
The Atmospheric Imaging Assembly (AIA) on {\it SDO} records 
full-disk images in seven EUV, two UV, and one white-light channel, 
with 0{\farcs}6 pixels and 10~s cadence.\footnote{Daily full-resolution 
images may be viewed at \texttt{http://sdowww.lmsal.com/suntoday}.}  
The Helioseismic and Magnetic Imager (HMI) provides longitudinal 
magnetograms with similar spatial resolution, taken every 45~s 
in \ion{Fe}{1} 617.3~nm.\footnote{See \texttt{http://hmi.stanford.edu}.}  
The AIA and HMI images were coaligned using the disk centers as the 
common reference point, after correcting for the factor of 0.844 
plate-scale difference, determined from the ratio of AIA 170.0~nm and 
HMI disk diameters in pixels.  The data employed here are all from 
the period 2010 August--September, early during the rising phase of 
solar cycle 24.

As an illustrative example, Figure 1 shows a large, decaying active region 
(NOAA 11100) in the southern hemisphere, as it appeared at 23:01~UT 
on 2010 August~20.  At the left are AIA images recorded in \ion{Fe}{9} 17.1~nm 
(characteristic temperature $T\sim 0.7$~MK), \ion{He}{2} 30.4~nm, and 
\ion{Fe}{14} 21.1~nm ($T\sim 2.0$~MK).  At the right are three versions 
of a simultaneous HMI magnetogram, saturated respectively at $\pm$100~G, 
$\pm$30~G (after smoothing to $2^{\prime\prime}$ resolution), and $\pm 0.1$~G 
(after smoothing with a $60^{\prime\prime}\times 60^{\prime\prime}$ 
running window).  From the 17.1~nm image, we see that the active region, 
which emerged one month earlier and is centered at latitude $L\sim -23^\circ$, 
is surrounded by a dark area that extends equatorward (poleward) as far as 
$L\sim -5^\circ$ ($L\sim -50^\circ$).  This canopy consists of 
leaf- or featherlike structures that fan out more or less radially 
from the active region.  The \ion{He}{2} 30.4~nm image shows a dark canopy 
of similar shape and extent; from this correspondence, we conclude that 
the dark 17.1~nm features are analogous to the \ion{He}{2} fibrils and 
filaments displayed in Wang (2001), which resemble but occupy larger 
volumes than their well-known H$\alpha$ counterparts.  (Unless otherwise 
noted, the term ``fibril'' hereinafter refers to features seen in 17.1~nm.)  
The canopy is much less visible in the \ion{Fe}{14} 21.1~nm image, being 
partially obscured by diffuse emission from the overlying coronal loops.  
Conversely, quiet Sun areas outside the canopy (as at the upper left and 
upper right corners of the images) are darker in the higher-temperature lines 
than in 17.1~nm.

Comparing the \ion{Fe}{9} image (Figure~1(a)) with the line-of-sight 
magnetograms (Figures~1(d) and 1(e)), we see that the dark features 
in 17.1~nm are generally centered above areas of {\it relatively} weak 
photospheric flux, as expected if they trace out horizontal fields.  
These locations of reduced $\vert B_r\vert$ include not only the 
supergranular cell interiors, but also more extensive areas of 
mixed polarity, where the spreading active-region flux encounters 
background regions of the opposite polarity.  The boxes in Figure~1 
highlight some of the dark fibrilar material that straddles these 
large-scale polarity inversion lines (PILs); the polarity distribution 
of the line-of-sight photospheric field, smoothed to the spatial scale 
of the supergranulation, is displayed in Figure~1(f).  Note that 
this dark canopy material is distinct from the arrowed filaments, 
which are closely aligned with the internal PIL of the active region.

Figure 2 shows a close-up of the northern edge of the active region canopy, 
as observed in \ion{Fe}{9} 17.1~nm at 17:01 and 23:01~UT on August~20.  
The corresponding line-of-sight magnetograms, smoothed to a resolution 
of 2$^{\prime\prime}$ (to reduce the noise level) and saturated at $\pm$30~G, 
are displayed at the right.  Arrows indicate the presumed local direction 
of the horizontal fibril fields, with the arrowheads pointing toward 
the negative-polarity footpoints.  The outer boundary of the dark canopy 
coincides very roughly with a string of positive-polarity network elements 
extending from northeast to southwest across the field of view; 
inside this ``front line,'' negative-polarity flux spreading outward from 
the decaying active region dominates.  Even here, however, small 
positive-polarity flux elements are almost everywhere present.  
Thus, it is often difficult to ascertain whether a given fibril structure 
is rooted in positive or negative polarity.

The boxed areas labeled ``A'' and ``B'' in Figure 2 enclose newly formed 
(or newly darkened) fibril structures located next to opposite-polarity 
flux elements in close contact, which in turn appear as compact 17.1~nm 
bright points.  By constructing time-lapse movies at 30-min 
cadence,\footnote{Reduced-resolution HMI and AIA movies can be viewed at 
\texttt{http://sdo.gsfc.nasa.gov/data/aiahmi/browse.php}.} we have verified 
that these flux elements are in the process of converging and canceling.  
The box labeled ``C'' encloses a string of small emerging bipoles; 
note, however, that most of the mixed-polarity flux in these magnetograms 
represents preexisting network and intranetwork elements that are 
undergoing random encounters in the supergranular flow field, rather than 
ephemeral regions (for a discussion of network, intranetwork, and 
ephemeral region fields and their interactions, see Martin 1988).

During the first 3 weeks of 2010 August, a dark canopy that stretched 
across $\sim$180$^\circ$ of longitude, extended from below the equator to 
above $L\sim +40^\circ$, and contained a number of new-cycle active regions, 
was observed rotating across the solar disk.  The left column of 
Figure~3 displays this giant northern-hemisphere canopy as it appeared on 
August~11 in an H$\alpha$ filtergram taken at the Big Bear Solar Observatory 
(BBSO), in a 17.1~nm image from AIA, and in an HMI magnetogram saturated 
at $\pm$30~G (after $2^{\prime\prime}\times 2^{\prime\prime}$ smoothing) 
and at $\pm 0.1$~G (after $60^{\prime\prime}\times 60^{\prime\prime}$ 
smoothing).  The right column of Figure~3 shows the same area one rotation 
later on September~8, after the active region fields have undergone further 
weakening and dispersal.  It is evident that the 17.1~nm fibrils are now more 
concentrated around the large-scale polarity inversions, forming structures 
that more closely resemble PIL-aligned filaments.  In particular, 
the northern section of the canopy has evolved into a U-shaped filament 
channel (see the boxed area) enclosing the sheared, negative-polarity remnant 
of the large active region on the southwest side of the canopy.  
Correspondingly, the H$\alpha$ image shows filament material extending 
along the northern edge of the canopy, which was not present on August~11.

Figure 4 focuses on the far western edge of the same northern-hemisphere 
canopy, as it appeared at 12:01~UT on August~8.  The dark fibrils 
originating from the negative-polarity plage around the sunspot 
(located at the bottom left corner of the images) are seen to fan out 
more or less radially into the nearby network; the outer endpoints can be 
presumed to have positive polarity, even though they are sometimes located 
where both polarities are in close proximity or where the flux is very weak.  
Somewhat farther to the west, where the dominant polarity changes 
from negative to positive, the fibril structures become oriented 
more or less parallel to the large-scale PIL.  Considering now the 
small active region on the right-hand side of these images, we observe 
a fountain pattern of dark fibrils that occupy a corridor of 
weak line-of-sight field and connect the negative-polarity plage to 
the positive-polarity background network lying to the west.  Viewed from 
the positive-polarity side of the PIL, the field lines point to the right, 
consistent with the ``dextral'' handedness characterizing the majority 
of northern-hemisphere filaments (Martin et al. 1994; Martin 1998).  
The fibrils and ``proto-filaments'' surrounding the large active region 
on the left-hand side of the images likewise exhibit dextral handedness.

The sequence of 17.1 nm images and magnetograms in Figure 5 shows how 
the northern edge of the giant canopy evolves during August~10--13 
(the field of view lies within the boxed area in Figures~3(a)--3(d)).  
Here, we see a collection of dark, featherlike structures lying between the 
predominantly positive-polarity area to the north and the sheared band of 
negative-polarity flux originating from the active region to the southwest.  
Early on August~10 (Figures~5(a) and 5(b)), the fibrils inside the 
white box are oriented almost perpendicular to the large-scale PIL, 
which is clearly defined by the line of strong negative-polarity 
network elements stretching from northeast to southwest.  Over the 
next two days, the negative-polarity flux spreads northward into the 
positive-polarity background region, and the two polarities intermingle; 
at the same time, the fibrils begin to bend in the direction of the PIL.  
In time-lapse movies, separate clumps of 17.1~nm fibrils oriented 
at different angles to each other are seen merging into longer structures 
with intermediate orientations (as in Figures~2 and 3 of 
Wang \& Muglach 2007).  Movies made from the HMI magnetograms show 
rapid flux cancellation occurring at the PIL, as well as rotational shearing, 
with the positive-polarity flux elements drifting eastward relative to 
the negative-polarity region to the south.  By August~13, the fibrils 
have coalesced into long, filament-like structures that are more or less 
aligned with the PIL.  It should be emphasized that the observed change 
in orientation of the fibrils relative to the PIL cannot be due to 
the photospheric differential rotation alone, which would cause a 
north--south aligned fibril with endpoints at latitudes 36$^\circ$ and 
42$^\circ$ to tilt only $\sim$14$^\circ$ over a 4-day period.  

The characteristic height or vertical extent $h_{171}$ of a 17.1~nm fibril 
above the photosphere can be estimated by comparing the corrugated texture 
of the limb in 17.1~nm with AIA images taken at 450.0~nm.  From the fact 
that the fibril structures typically protrude 
$\sim$6$^{\prime\prime}$--8$^{\prime\prime}$ beyond the white-light limb, 
we deduce that $h_{171}\sim 4000$--6000~km, well above the nominal height 
of the chromospheric--coronal transition region.

\section{Physical Interpretation}
We proceed from the basic premise that the dark features seen in 
\ion{Fe}{9} 17.1~nm represent horizontal flux tubes that contain 
chromospheric material and connect photospheric flux elements of 
opposite polarity.  (The cool material may consist of neutral hydrogen 
and helium which absorb the EUV radiation impinging from below: see, e.g., 
Chiuderi~Drago et al. 2001; Rutten 2007.)  Such horizontal fields form 
low-lying canopies that fan out from the edges of magnetic flux concentrations, 
whether they be active regions, sunspots, or network boundaries/vertices 
(Harvey 2005, 2006).  In emerging active regions, the flux balloons outward 
in a dipole-like configuration and the surface-skimming field lines become 
connected to background network and intranetwork elements, forming a 
pattern of dark fibrils diverging from the area occupied by strong plage 
(as sketched in Figure~10 of Wang \& Muglach 2007).  

We expect the fibril pattern to continue to expand outward even after 
the active region has fully emerged, because of the diffusive effect of 
the nonsteady supergranular convection, which causes the plage to 
disintegrate and spread into the weaker-field background.  The flux 
is swept to the boundaries of the randomly distributed supergranular cells, 
which have a characteristic diameter of $\sim$30,000~km and lifetime 
of 1--2~days.  As the cells decay and re-form at other locations, 
the fibril patterns continually change in response, in such a way that 
the fibrils remain centered above areas of relatively weak photospheric field, 
while their endpoints remain anchored in network or intranetwork elements 
of opposite polarity.  Flux cancellation will act to weaken the 
network fields, with the fibrils rooted in the canceling flux merging 
into longer structures.  Other fibrils may split apart if the 
convective flows bring sufficiently strong network or intranetwork elements 
between their endpoints or if new bipoles emerge underneath them.

The canopy grows until the local rate of flux cancellation becomes so high 
that the diffusing active-region flux can no longer penetrate into the 
surrounding network.  The outer boundary of the canopy thus stabilizes 
around a large-scale PIL separating the active region remnant from 
background field dominated by the opposite polarity.  Continued 
flux cancellation around the PIL acts to submerge the transverse field 
component; as a result, the fibrils in this region become increasingly 
aligned with the PIL, and eventually coalesce into long filaments 
(see, e.g., van Ballegooijen \& Martens 1989).  This surviving axial field 
was generated in part by the photospheric differential rotation, 
which also acts to speed up the diffusive annihilation of flux at the PIL; 
in addition, it may contain a contribution arising from the 
intrinsic helicity of the original active region.  However, shearing motions 
alone cannot account for changes in the orientation of the fibrils 
on timescales of a few days; flux cancellation is required to produce 
the observed rapid alignment with the PIL (cf. Wang \& Muglach 2007).

An important observational result is that the dark 17.1 nm fibrils, 
once they have become detached via reconnection and diffusive transport 
from their active region source, tend to accumulate in areas of weak or 
mixed-polarity flux.  This tendency is easily understood, given our 
assumption that fibrils are low-lying horizontal fields.  Such field lines 
are more likely to reconnect with other low loops than with the high loops 
that are rooted inside strong unipolar flux concentrations (see Figure~6).  
Small, low-lying loops are in turn associated with mixed-polarity flux, 
whether in the form of canceling network elements, intranetwork fields, 
or small ephemeral regions.  The net effect is to concentrate the fibrils 
around the large-scale background PILs surrounding the active region, 
where the bulk of the mixed-polarity flux is located.  Systematic flux 
cancellation in this region then expedites the conversion of the fibrils 
into PIL-aligned filaments.

In his seminal studies of H$\alpha$ fine structure, Foukal (1971a, 1971b) 
came to the conclusion that H$\alpha$ fibrils cannot connect across 
supergranule cells, but must be open-ended absorbing features that bend 
upward and link to remote areas where the network has the opposite polarity.  
This conclusion was based on inspection of a then-available magnetogram 
showing that the opposite sides of a supergranule are usually of the same 
polarity.  However, it has since become clear from higher-resolution 
observations that so-called unipolar regions contain large amounts of 
minority-polarity flux; indeed, Schrijver \& Title (2003) assert that 
as much as one-half of the strong network flux connects down into the 
immediately surrounding intranetwork areas.  We also note that, because 
the magnetic network becomes increasingly fragmented toward the peripheries 
of active regions, horizontal field lines that are not ``captured'' by 
nearby loops are as likely to be channeled through gaps in the network 
as to be deflected sharply upward into the corona, as in Foukal's picture.  

High-cadence movies made from the full-resolution 17.1~nm images suggest 
that the fibril material streams continually from one footpoint to the other.  
These flows may be triggered by reconnection between the small loops 
and the fibril fields, generating chromospheric jets that inject new material 
into the fibrils.  A similar process is thought to supply mass to H$\alpha$ 
and \ion{He}{2} 30.4~nm filaments, where upflows and downflows of 
$\sim$5--70 ~km~s$^{-1}$ are observed (Zirker et al. 1998; Litvinenko \& 
Martin 1999; Wang 1999; Kucera et al. 2003).

We have earlier remarked that areas of weak, ``salt-and-pepper'' fields 
far from active regions (such as that at the upper right corner of the images 
in Figure~1) are brighter than the canopy regions in \ion{Fe}{9} 17.1~nm, 
whereas the opposite is the case in higher-temperature coronal lines.  
The darkness of the 17.1~nm canopy is due to the presence of the organized, 
large-scale horizontal fields originating from the active region; 
in the quiet Sun, the fibrils are smaller and oriented more randomly, 
except around large-scale PILs, where systematic flux cancellation 
takes place.  In contrast, the emission from coronal loops is correlated 
with their footpoint field strengths, so the quiet Sun generally appears 
darker than the areas immediately surrounding active regions when observed 
in lines such as \ion{Fe}{12} 19.3~nm and \ion{Fe}{14} 21.1~nm.

\section{Conclusions}
We have used high-resolution \ion{Fe}{9} 17.1~nm images and longitudinal 
magnetograms from {\it SDO} to explore the relationship between the 
circumfacular areas or ``dark canopies'' surrounding active regions and 
the evolving photospheric field.  Our main conclusions may be summarized 
as follows:

1. As has long been recognized, the canopies consist of dark fibril-like 
structures that appear as soon as the active regions emerge; 
the low-lying horizontal field lines balloon out over the surrounding area 
and reconnect with the background network to form a quasiradial or 
vortical pattern.  The 17.1~nm fibrils overlie areas of relatively weak 
photospheric field.

2. The nonsteady supergranular convection causes the fibril fields 
to spread outward from the active region, with the outer boundary 
of the dark canopy stabilizing where the diffusing flux encounters a 
unipolar region of the opposite sign.

3. The 17.1~nm fibrils are often rooted in mixed-polarity regions, 
because of the tendency for the horizontal flux tubes to reconnect with 
small, low-lying loops.

4. As a result of this attraction toward mixed-polarity flux, 
the diffusing 17.1~nm fibrils gradually accumulate around large-scale PILs.

5. Systematic flux cancellation at the PIL removes the component of 
the field transverse to the PIL and progressively converts the fibrils 
into PIL-aligned structures.

6. In the absence of new flux emergence, active region canopies thus 
evolve toward a state where the dark material becomes concentrated 
around the surrounding background PILs, forming proto-filaments and 
filaments.

This study, like that of Wang \& Muglach (2007) where H$\alpha$ 
observations were employed, points to the primary role of photospheric 
flux cancellation and fieldline reconnection in the evolution of fibrils 
into filaments.  By making full use of the high temporal resolution 
of the {\it SDO} observations, which we have not properly exploited here, 
it should be possible to track the evolution of individual fibril structures, 
to characterize their mass flows, to pinpoint their footpoint locations, 
to clarify the relationship between fibrils and chromospheric jets, 
and to determine more precisely how flux cancellation leads to the 
coalescence of fibrils and their transformation into PIL-aligned 
filament channels and filaments.

\acknowledgments
We thank N. R. Sheeley, Jr., P. R. Young, and P. Foukal for stimulating 
discussions, and the referee for detailed and constructive criticism 
of an earlier version of the manuscript.  The data employed here were provided 
courtesy of NASA/{\it SDO}, the AIA and HMI science teams, and BBSO/NJIT.  
This work was supported by NASA, NSF, and the Office of Naval Research.

\clearpage
\begin{figure*}
\vspace{-3cm}
\centerline{\includegraphics[width=45pc]{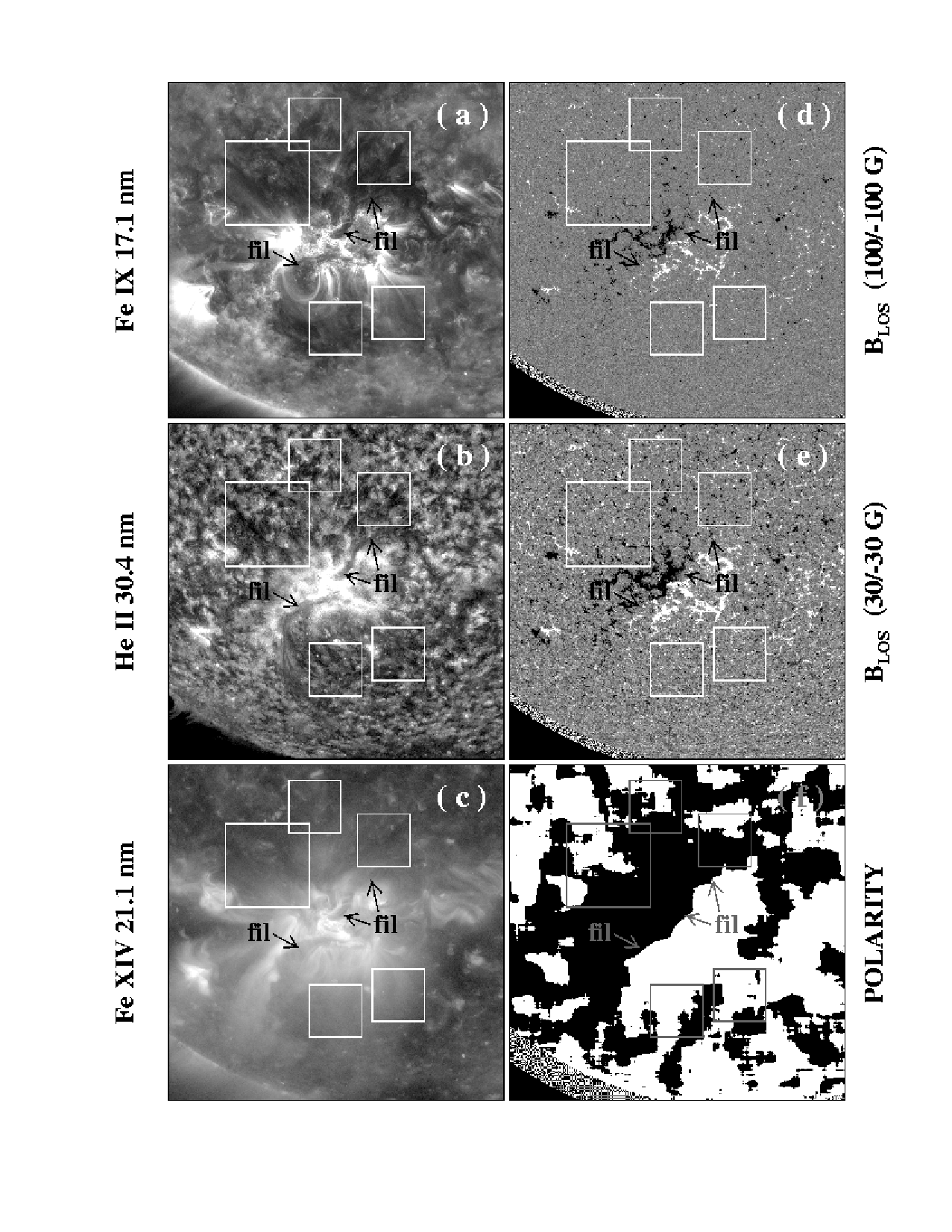}}
\vspace{-2.5cm}
\caption{AIA and HMI images showing a ``dark canopy'' surrounding 
an active region in the southern hemisphere, 2010 August~20 at 23:01~UT.  
(a) \ion{Fe}{9} 17.1~nm.  (b) \ion{He}{2} 30.4~nm.  (c) \ion{Fe}{14} 21.1~nm.  
(d) Simultaneous line-of-sight magnetogram saturated at $\pm$100~G.  
(e) The same magnetogram saturated at $\pm$30~G after 
$2^{\prime\prime}\times 2^{\prime\prime}$ smoothing.  (f) Polarity 
distribution after smoothing the magnetogram with a 
$60^{\prime\prime}\times 60^{\prime\prime}$ running window.  Boxes 
highlight areas containing dark fibrilar material and relatively weak 
longitudinal field, where the spreading active-region flux encounters 
background network of the opposite polarity.  Arrows point to a large 
filament aligned with the internal PIL of the active region.}
\end{figure*}

\clearpage
\begin{figure*}
\vspace{-3cm}
\centerline{\includegraphics[width=45pc]{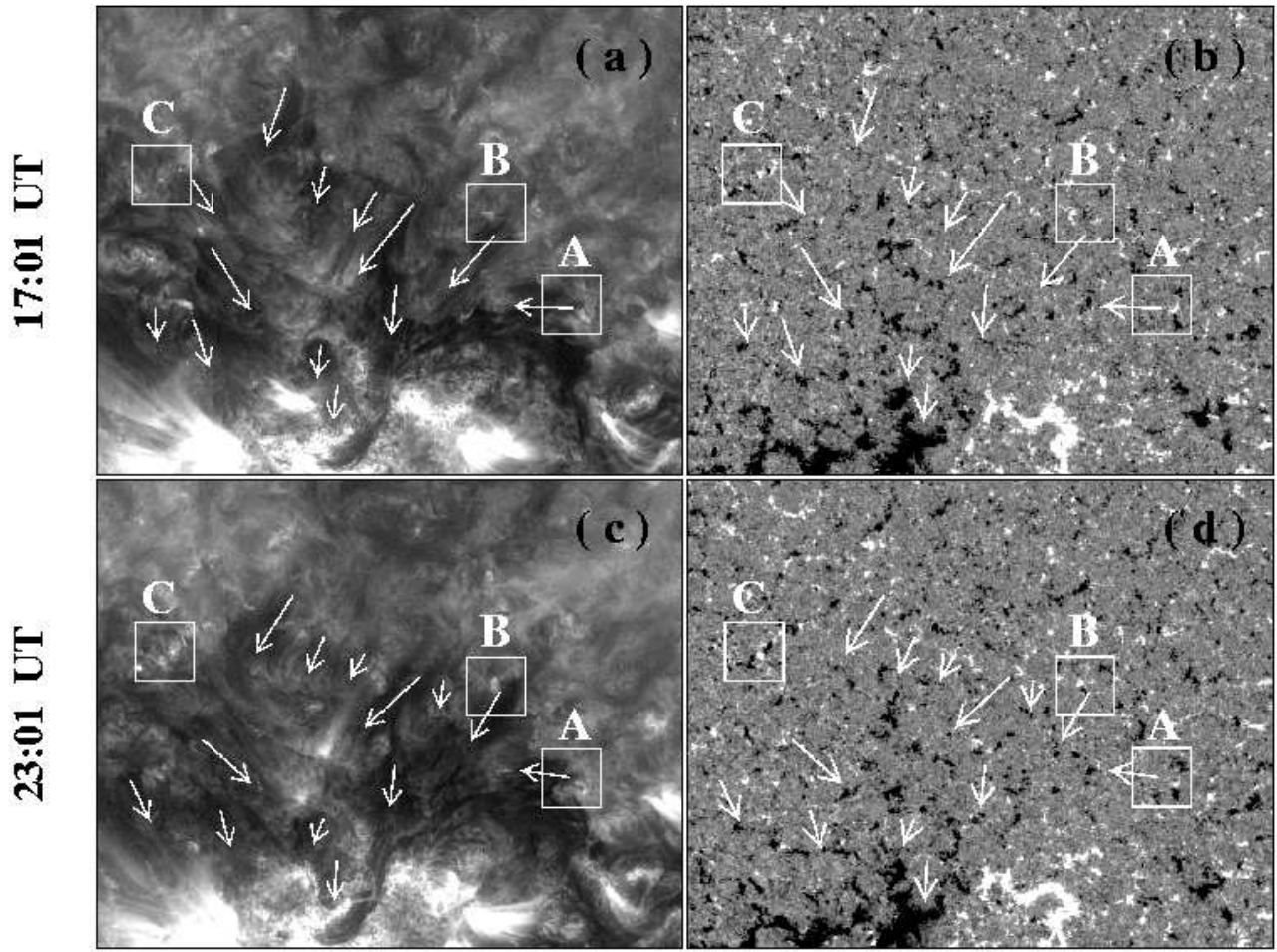}}
\vspace{-5cm}
\caption{Close-up of the northern edge of the active region canopy 
in Figure~1.  Field of view has dimensions 
$530^{\prime\prime}\times 424^{\prime\prime}$; north is up and west 
is to the right.  (a) \ion{Fe}{9} 17.1~nm image recorded at 17:01~UT 
on August~20.  (b) Simultaneous line-of-sight magnetogram, smoothed to 
a resolution of 2$^{\prime\prime}$ and saturated at $\pm$30~G.  
(c) \ion{Fe}{9} 17.1~nm image recorded at 23:01~UT on August~20.  
(d) Simultaneous line-of-sight magnetogram.  Arrows indicate the 
presumed local direction of the horizontal fibril fields, with the 
arrowheads pointing toward the negative-polarity footpoints.  Boxes 
labeled ``A'' and ``B'' highlight canceling magnetic flux elements; 
box ``C'' encloses newly emerged ephemeral regions.}
\end{figure*}

\clearpage
\begin{figure*}
\vspace{-4cm}
\centerline{\includegraphics[width=45pc]{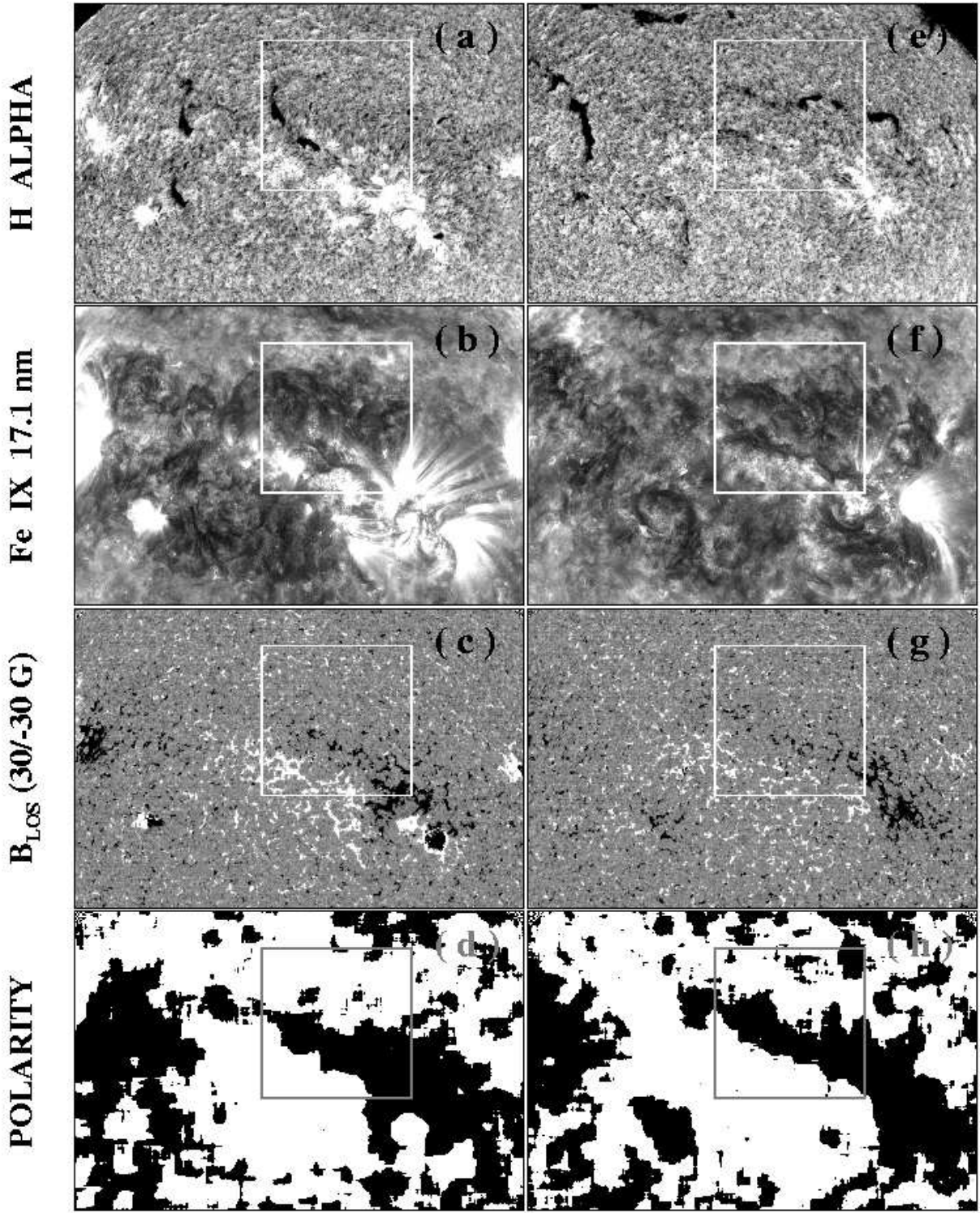}}
\vspace{-2cm}
\caption{Giant northern-hemisphere canopy observed on 2010 August~11 
(left panels) and one rotation later on September~8 (right panels).  
Field of view has dimensions $1268^{\prime\prime}\times 845^{\prime\prime}$.  
(a) BBSO H$\alpha$ filtergram taken at 16:03~UT on August~11 (line-center 
observations with 0.025~nm bandpass).  (b) AIA \ion{Fe}{9} 17.1~nm image 
recorded at 23:01~UT on August~11.  (c) HMI longitudinal magnetogram 
(23:01~UT), saturated at $\pm$30~G after 
$2^{\prime\prime}\times 2^{\prime\prime}$ smoothing.  (d) Corresponding 
polarity distribution after $60^{\prime\prime}\times 60^{\prime\prime}$ 
smoothing.  (e) BBSO H$\alpha$ filtergram taken at 16:07~UT on September~8.  
(f) 17.1~nm image recorded at 05:01~UT on September~8.  (g) longitudinal 
magnetogram recorded at 05:01~UT.  (h) Corresponding polarity distribution.  
The boxed area evolves into a U-shaped filament channel as the active region 
fields decay and the dark 17.1~nm fibrils become increasingly concentrated 
around the large-scale PILs.}
\end{figure*}

\clearpage
\begin{figure*}
\vspace{-1cm}
\centerline{\includegraphics[width=40pc]{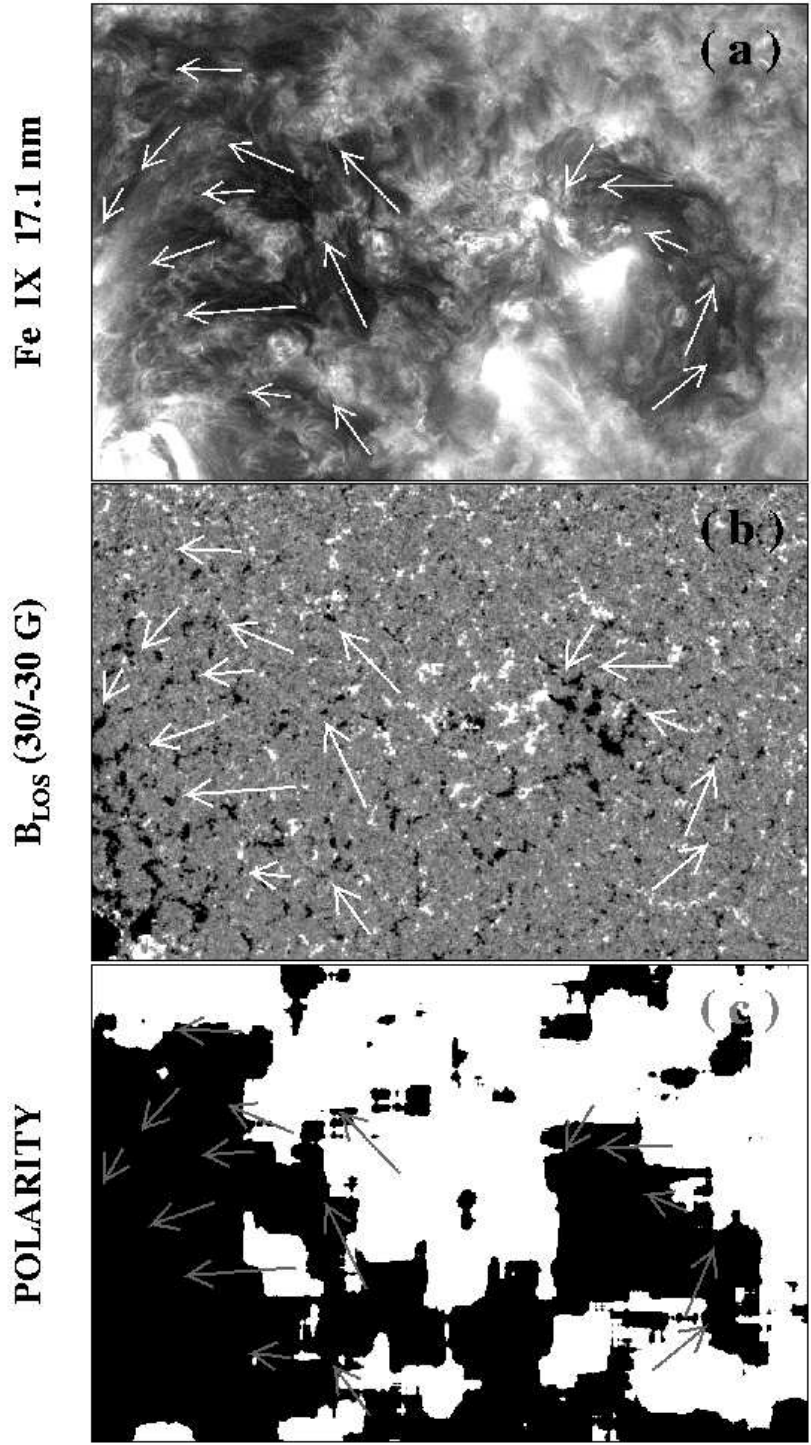}}
\vspace{-1cm}
\caption{Far western edge of the giant northern-hemisphere canopy of Figure~3, 
as it appeared at 12:01~UT on August~8.  Field of view has dimensions 
$634^{\prime\prime}\times 423^{\prime\prime}$.  (a) \ion{Fe}{9} 17.1~nm 
image.  (b) Corresponding line-of-sight magnetogram, saturated at $\pm$30~G 
after $2^{\prime\prime}\times 2^{\prime\prime}$ smoothing.  (c) Polarity 
distribution after $60^{\prime\prime}\times 60^{\prime\prime}$ smoothing.  
Arrows indicate the presumed local direction of the horizontal fibril fields.}
\end{figure*}

\clearpage
\begin{figure*}
\vspace{-2.5cm}
\centerline{\includegraphics[width=45pc]{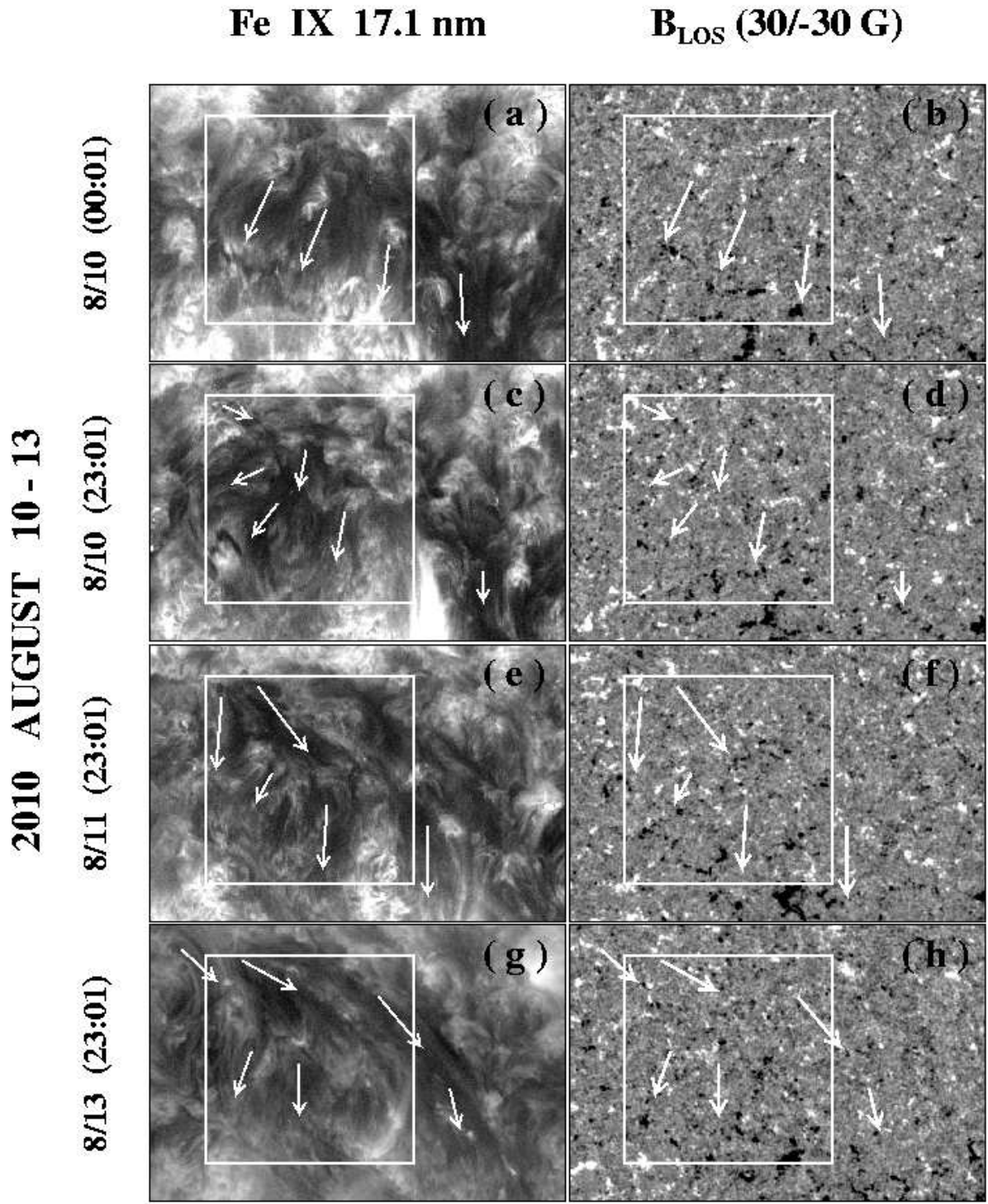}}
\vspace{-3cm}
\caption{Sequence of 17.1~nm images (left) and corresponding longitudinal 
magnetograms (right), showing the evolution of the northern edge of the 
giant canopy of Figure~3 during August~10--13.  Note the change in the 
orientation of the fibril structures inside the 
$158^{\prime\prime}\times 158^{\prime\prime}$ boxed area (centered at 
$L\sim +37^\circ$), from nearly perpendicular to nearly parallel to the 
large-scale PIL, as the negative-polarity flux from the sheared active-region 
remnant diffuses into the positive-polarity background region to the north.  
Arrows indicate the presumed local direction of the horizontal fibril fields.  
The line-of-sight magnetograms are again saturated at $\pm$30~G after 
$2^{\prime\prime}\times 2^{\prime\prime}$ smoothing.}
\end{figure*}

\clearpage
\begin{figure*}
\vspace{1cm}
\centerline{\includegraphics[width=30pc]{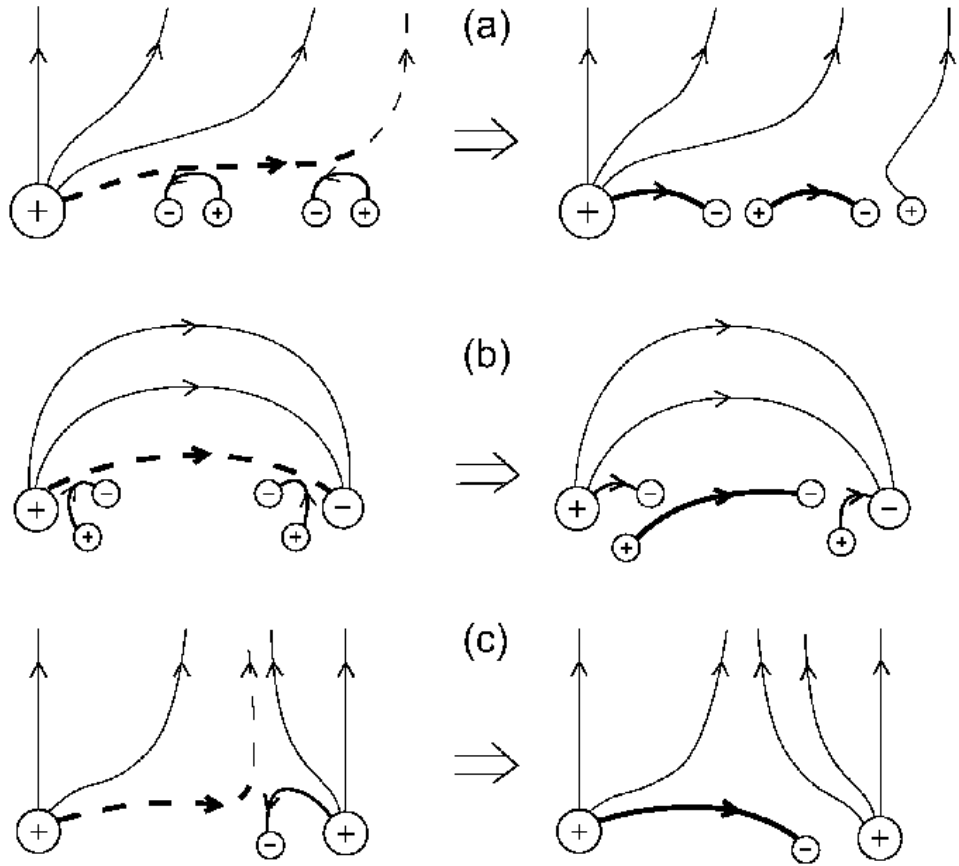}}
\vspace{1.5cm}
\caption{Low-lying, horizontal field lines have a natural tendency 
to become linked to weak, mixed-polarity network and intranetwork flux.  
(a) A long fibril (dashed) anchored in active region plage and extending out 
into the background network reconnects with a pair of small bipoles 
and splits into two pieces.  (b) A fibril that initially links two strong, 
opposite-polarity network elements reconnects with the surrounding 
small bipoles and becomes rooted in weak, mixed-polarity flux.  
(c) Reconnection with small bipoles may also give rise to fibrils whose 
endpoints are located near strong network elements of the same polarity.  
In all of these cases, reconnection occurs at low heights and may trigger 
chromospheric jets that supply mass to the fibrils.}
\end{figure*}

\end{document}